\documentclass[10pt]{article}

\usepackage{authblk}
\usepackage{natbib}
\usepackage{mathtools}
\usepackage{amsmath}
\usepackage{amsthm}
\usepackage{amssymb}
\usepackage{amsbsy}
\usepackage{amsfonts}
\usepackage{amscd}
\usepackage{mathrsfs}
\usepackage{bm}
\usepackage{breqn}
\usepackage{color, soul}
\usepackage{enumerate}
\usepackage{empheq}
\usepackage{rotating}
\usepackage{ftnxtra}
\usepackage{fnpos}
\usepackage[titletoc,toc,title]{appendix}
\usepackage{euscript}
\usepackage{graphicx}
\usepackage{epsfig}
\usepackage{epstopdf}
\DeclareGraphicsExtensions{.pdf,.png,.jpg,.eps}
\usepackage{pstool}
\usepackage{upgreek}
\usepackage{mathrsfs}

\usepackage{psfrag}

\numberwithin{equation}{section}

\theoremstyle{plain}	
\newtheorem{thm}{Theorem}[section]

\newtheorem{prop}[thm]{Proposition}
\newtheorem*{prop*}{Proposition} 
\theoremstyle{definition}	

\newtheorem{remark}[thm]{Remark}

\usepackage{graphicx}
\usepackage{lipsum}

\usepackage{caption}
\usepackage{subcaption}

\usepackage{hyperref}
\hypersetup{colorlinks=true, linkcolor=blue}
\hypersetup{colorlinks=true,citecolor=blue}

\DeclareMathAlphabet{\mathpzc}{OT1}{pzc}{m}{it}

\usepackage{amsmath, amsthm, amssymb}

\usepackage{cleveref}

\DeclarePairedDelimiter\abs{\lvert}{\rvert}

\makeatletter
\newsavebox{\@brx}
\newcommand{\llangle}[1][]{\savebox{\@brx}{\(\m@th{#1\langle}\)}%
  \mathopen{\copy\@brx\mkern2mu\kern-0.9\wd\@brx\usebox{\@brx}}}
\newcommand{\rrangle}[1][]{\savebox{\@brx}{\(\m@th{#1\rangle}\)}%
  \mathclose{\copy\@brx\mkern2mu\kern-0.9\wd\@brx\usebox{\@brx}}}%
\let\oldabs\abs
\def\abs{\@ifstar{\oldabs}{\oldabs*}}
\makeatother

\usepackage{accents}

    {\end{bmatrix}}%

\usepackage{enumitem}

\usepackage[utf8]{inputenc}
\usepackage{dutchcal}
\usepackage{multirow}

\usepackage{xcolor}

\usepackage{setspace}

\begin{document}

\title{\textbf{Universal Deformations and Inhomogeneities in Isotropic Cauchy Elasticity}}

\author[1,2]{Arash Yavari\thanks{Corresponding author, e-mail: arash.yavari@ce.gatech.edu}}
\affil[1]{\small \textit{School of Civil and Environmental Engineering, Georgia Institute of Technology, Atlanta, GA 30332, USA}}
\affil[2]{\small \textit{The George W. Woodruff School of Mechanical Engineering, Georgia Institute of Technology, Atlanta, GA 30332, USA}}

\maketitle

\vskip 0.1in
{\centering Dedicated to Professor Kaushik Bhattacharya on the occasion of his $60$th birthday. \par}

\begin{abstract}
For a given class of materials, \emph{universal deformations} are those deformations that can be maintained in the absence of body forces and by applying solely boundary tractions. For inhomogeneous bodies, in addition to the universality constraints that determine the universal deformations, there are extra constraints on the form of the material inhomogeneities---\emph{universal inhomogeneity constraints}. Those inhomogeneities compatible with the universal inhomogeneity constraints are called \emph{universal inhomogeneities}. In a Cauchy elastic solid, stress at a given point and at an instance of time is a function of strain at that point and that exact moment in time, without any dependence on prior history.  A Cauchy elastic solid does not necessarily have an energy function, i.e., Cauchy elastic solids are, in general, non-hyperelastic (or non-Green elastic). In this paper we characterize universal deformations in both compressible and incompressible inhomogeneous isotropic Cauchy elasticity. As Cauchy elasticity includes hyperelasticity, one expects the universal deformations of Cauchy elasticity to be a subset of those of hyperelasticity both in the compressible and incompressible cases. It is also expected that the universal inhomogeneity constraints to be more stringent than those of hyperelasticity, and hence, the set of universal inhomogeneities to be smaller than that of hyperelasticity. We prove the somewhat unexpected result that the sets of universal deformations of isotropic Cauchy elasticity and isotropic hyperelasticity are identical, in both the compressible and incompressible cases. We also prove that their corresponding universal inhomogeneities are identical as well. 
\end{abstract}

\begin{description}
\item[Keywords:] Universal deformation, universal inhomogeneity, Cauchy elasticity, hyperelasticity, Green elasticity, isotropic solids.
\end{description}


\section{Introduction}

Within a given class of materials, universal deformations refer to those deformations that can be maintained in the absence of body forces and by applying only boundary tractions, for any member of the material class. 
Universal deformations do not depend on the particular material within the class. However, the boundary tractions necessary to sustain a universal deformation depend on the specific material.
Universal deformations have played a crucial role in nonlinear elasticity and anelasticity (in the sense of \citet{Eckart1948}): i) They have had an important organizational role in the semi-inverse solutions in nonlinear elasticity \citep{Knowles1979,Polignone1991,DePascalis2009,Tadmor2012,Goriely2017}, and more recently in anelasticity \citep{KumarYavari2023} and viscoelasticity \citep{SadikYavari2024}. 
ii) They offer guidance for designing experiments for determining the constitutive relations of a specific material \citep{Rivlin1951,Saccomandi2001}.\footnote{The following quote from \citep[p.~89]{DoyleEricksen1956} explains it best: ``From the standpoint of attempting to determine the form of $\Sigma$ for a given material by comparing general solutions with results of experiment, it appears that the solutions which are most useful are those which correspond to deformations which can be produced in every material of the type considered by the application of surface tractions only." Here, $\Sigma$ is the energy function and by ``general" they mean ``universal".}
iii) All the existing exact solutions of defects in nonlinear solids are related to universal deformations \citep{Wesolowski1968,Gairola1979,Zubov1997,YavariGoriely2012a,YavariGoriely2012b,YavariGoriely2013a,YavariGoriely2014,Golgoon2018}.
iv) Universal deformations have been important for finding exact solutions for distributed finite eigenstrains in nonlinear solids and the nonlinear analogues of Eshelby's inclusion problem \citep{YavariGoriely2013b,YavariGoriely2015,Golgoon2017,Yavari2021Eshelby}.
v) These exact solutions have been used as benchmark problems in computational mechanics \citep{Dragoni1996,Saccomandi2001,Chi2015,Shojaei2018}.
vi) Universal deformations have been used in deriving effective properties for nonlinear composites \citep{Hashin1985,Lopez2012,Golgoon2021}.

The systematic study of universal deformations began in the 1950s by Jerry Ericksen \citep{Ericksen1954,Ericksen1955} for homogeneous compressible and incompressible isotropic solids.
His work was influenced by the earlier contributions of Ronald Rivlin \citep{Rivlin1948, Rivlin1949a, Rivlin1949b}. 
\citet{Ericksen1955} proved that in homogeneous compressible isotropic solids, universal deformations are homogeneous.
The characterization of universal deformations in the presence of internal constraints is a particularly challenging problem \citep{Saccomandi2001}.  \citet{Ericksen1954} discovered four families of universal deformations for incompressible isotropic elastic solids. Initially, he speculated that a deformation with constant principal invariants is homogeneous, but this conjecture was proven incorrect \citep{Fosdick1966}.
Subsequently, a fifth family of universal deformations was found \citep{SinghPipkin1965,KlingbeilShield1966}.
The fifth family of universal deformations have constant principal invariants but are not homogeneous. As of now, it is not known whether there exist additional inhomogeneous constant-principal invariant universal deformations (Ericksen's problem).
\noindent The following are the six known families of universal deformations:
\begin{itemize}[topsep=2pt,noitemsep, leftmargin=10pt]
\item Family 0: Homogeneous deformations
\item Family 1: Bending, stretching, and shearing of a rectangular block
\item Family 2: Straightening, stretching, and shearing of a sector of a cylindrical shell
\item Family 3: Inflation, bending, torsion, extension, and shearing of a sector of an annular wedge
\item Family 4: Inflation/inversion of a sector of a spherical shell
\item Family 5: Inflation, bending, extension, and azimuthal shearing of an annular wedge
\end{itemize}

\vskip 0.02in
There have been several attempts to solve Ericksen's problem in the past few decades \citep{Marris1970,Kafadar1972,Marris1975,Marris1982}.
\citet{Fosdick1969} showed that for the case of plane deformations with uniform transverse stretch there are no new solutions other than the known families.
\citet{Fosdick1971} reached the same conclusion for radially-symmetric deformations.

In a \emph{simple material}, stress at any given point depends only on the history of the deformation gradient at that point up to time $t$ \citep{Noll1958}. \citet{Carroll1967} demonstrated  that the above known universal deformations of homogenous incompressible isotropic elastic solids are universal for simple materials as well.

The study of universal deformations was extended to inhomogeneous anisotropic solids in \citep{YavariGoriely2021,Yavari2021,YavariGoriely2023Universal}.\footnote{Prior to our work, there had been some limited work on universal deformations in anisotropic solids \citep{Ericksen1954Anisotropic}.} 
These comprehensive studies included both compressible and incompressible isotropic, transversely isotropic, orthotropic, and monoclinic solids. 
For these three classes of compressible anisotropic solids, it was shown that universal deformations are homogenous, and the material preferred directions are uniform. Additionally, for isotropic solids and each of the three classes of anisotropic solids, the corresponding universal inhomogeneities---these represent inhomogeneities in the energy function that are compatible with the universality constraints---were characterized.
The corresponding universal inhomogeneities for each of the above six known families of universal deformations were determined for inhomogeneus isotropic and the three classes of inhomogeneous incompressible anisotropic solids.

In linear elasticity, universal displacements are the counterparts to universal deformations \citep{Truesdell1966,Gurtin1972,Carroll1973,Yavari2020}. \citet{Yavari2020} demonstrated the explicit dependence of universal displacements on the symmetry class of the material. Specifically, the larger the symmetry group, the larger the corresponding set of universal displacements. 
Therefore, isotropic solids have the largest set of universal displacements, while triclinic solids possess the smallest set of universal displacements. 
The investigation into universal displacements has also been extended to inhomogeneous solids \citep{YavariGoriely2022} and linear anelasticity \citep{Yavari2022Anelastic-Universality}.

A class of materials with internal constraints, significant in engineering applications, consists of materials reinforced with inextensible fibers \citep{AdkinsRivlin1955,Pipkin1971,Pipkin1974,Pipkin1979}. Despite their importance, there is a scarcity of literature on the universal deformations of fiber-reinforced solids. \citet{Beskos1972} studied homogeneous compressible isotropic solids reinforced with inextensible fibers, investigating whether universal deformations of incompressible isotropic solids are universal for this class as well. Families $1$, $2$, $3$, and $4$ were specifically analyzed, revealing that certain subsets of these families are universal for specific fiber distributions. Interestingly, all these universal deformations are found to be homogeneous except for the shearing of a circular tube with circumferential fibers. \citet{Beatty1978} further examined homogeneous compressible isotropic solids reinforced by a single family of inextensible fibers. He investigated the problem of finding all those fiber distributions for which homogenous deformations are universal. He proved that there are only three types of such fiber distributions, all characterized by straight fibers.
In a recent study, \citet{Yavari2023} studied universal displacements in compressible anisotropic linear elastic solids reinforced by a single family of inextensible fibers. For each symmetry class, and under the assumption of a uniform distribution of straight fibers respecting the corresponding symmetry, the respective universal displacements were characterized. Interestingly, it was observed that, except for triclinic and cubic solids, the presence of inextensible fibers enlarges the set of universal displacements within the other five classes.\footnote{It should be noted that a fiber-reinforced solid with straight fibers cannot be isotropic.}

In recent years, Ericksen's analysis has been extended to anelasticity. 
\citet{YavariGoriely2016} showed that in compressible anelasticity, universal deformations are covariantly homogeneous. 
Universal deformations and eigestrains in incompressible anelasticity were studied by \citet{Goodbrake2020}.
It was observed that the six known families of universal deformations are invariant under specific Lie subgroups of the special Euclidean group.
There are also some recent studies of universal deformations and eigenstrains in accreting bodies \citep{YavariPradhan2022,YavariAccretion2023,PradhanYavari2023}.
There have also been studies of universal deformations in liquid crystal elastomers \citep{LeeBhattacharya2023,MihaiGoriely2023}

In this paper, we extend the study of universal deformations and inhomogeneities to inhomogeneous compressible and incompressible isotropic Cauchy elasticity. Cauchy elastic solids may not necessarily have an energy function and include hyperelastic solids (Green elastic solids) as a special case. This suggests that Cauchy elasticity may have more stringent universality and universal inhomogeneity constraints compared to those of hyperelasticity. This, in turn, leads one to anticipate smaller sets of universal deformations and universal inhomogeneities for Cauchy elasticity compared to those of hyperelasticity. We prove the somewhat unexpected result that the universal deformations and inhomogeneities of Cauchy elasticity are identical to those of Green elasticity in both the compressible and incompressible cases.

This paper is organized as follows. In \S\ref{C-Elasticity}, Cauchy elasticity is briefly reviewed. Universal deformations of inhomogeneous compressible isotropic Cauchy elasticity are characterized in \S\ref{C-Elasticity-UC}. In \S\ref{C-Elasticity-UI}, the same problem is studied for inhomogeneous incompressible isotropic Cauchy elasticity. Conclusions are given in \S\ref{Sec:Conclusions}.

\section{Cauchy Elasticity} \label{C-Elasticity}

Let us consider a body that in its undeformed configuration is identified with an embedded submanifold $\mathcal{B}$ of the Euclidean ambient space $\mathcal{S}$. The flat metric of the Euclidean ambient space is denoted by $\mathbf{g}$ and the induced metric on the body in its reference configuration by $\mathbf{G}=\mathbf{g}|_{\mathcal{B}}$. Deformation is a map from $\mathcal{B}$ to the ambient space, i.e., $\varphi:\mathcal{B}\to\mathcal{C}\subset\mathcal{S}$, where $\mathcal{C}=\varphi(\mathcal{B})$ is the current configuration. The tangent map of $\varphi$ is the so-called deformation gradient $\mathbf{F}=T\varphi$ (a metric independent map), which at each material point $X \in \mathcal{B}$ is a linear map $\mathbf{F}(X):T_{X}\mathcal{B}\rightarrow T_{\varphi(X)}\mathcal{C}$. 
With respect to the coordinate charts $\{X^A\}$ and $\{x^a\}$ for $\mathcal{B}$ and $\mathcal{C}$, respectively, deformation gradient has components $F^a{}_A=\frac{\partial \varphi^a}{\partial X^A}$.
The transpose of deformation gradient $\mathbf{F}^{\textsf{T}}$ has components $(F^{\textsf{T}})^A{}_{a}=g_{ab}\,F^b{}_{B}\,G^{AB}$.
The right Cauchy-Green strain is defined as $\mathbf{C}=\mathbf{F}^{\textsf{T}}\mathbf{F}$ and has components $C^A{}_{B}=(F^{\textsf{T}})^A{}_{a}F^a{}_{B}$.
Thus, $C_{AB}=(g_{ab}\circ \varphi)\,F^a{}_{A}\,F^b{}_{B}$, which means that the right Cauchy-Green strain is  the pull-back of the spatial metric to the reference configuration, i.e., $\mathbf{C}^\flat=\varphi^*\mathbf{g}$, where $\flat$ is the flat operator induced by the metric $\mathbf{G}$ (which lowers indices).
The left Cauchy-Green strain is defined as $\mathbf{B}^{\sharp}=\varphi^*(\mathbf{g}^{\sharp})$, which has components $B^{AB}=F^{-A}{}_a\,F^{-B}{}_b\,g^{ab}$. 
The spatial analogue of $\mathbf{C}^\flat$ is defined as $\mathbf{c}^\flat=\varphi_*\mathbf{G}$, which has components $c_{ab}=F^{-A}{}_a\,F^{-B}{}_b\,G_{AB}$.
Similarly, the spatial analogue of $\mathbf{B}^{\sharp}$ is $\mathbf{b}^{\sharp}=\varphi_*(\mathbf{G}^{\sharp})$, which has components $b^{ab}=F^a{}_{A}\,F^b{}_{B}\,G^{AB}$.
Recall that $\mathbf{b}=\mathbf{c}^{-1}$.
The two tensors $\mathbf{C}$ and $\mathbf{b}$ have the same principal invariants $I_1$, $I_2$, and $I_3$, which are defined as \citep{Ogden1984,MarsdenHughes1994} $I_1 =\operatorname{tr}\mathbf{b}=b^{ab}\,g_{ab}$, $I_2=\frac{1}{2}\left(I_1^2-\operatorname{tr}\mathbf{b}^2\right)=\frac{1}{2}\left(I_1^2-b^{ab}b^{cd}\,g_{ac}\,g_{bd}\right)$, and $I_3=\det \mathbf{b}$.

In Cauchy elasticity, stress at a point and a given moment in time is explicitly a function of strain at that point and that particular moment in time \citep{Cauchy1828,Truesdell1952,TruesdellNoll2004}. However, an energy function does not necessarily exist.\footnote{It is important to note that Cauchy elasticity does not encompass all elastic solids. In recent years, there has been some interest in implicit constitutive equations, e.g., constitutive equations of the form $\boldsymbol{\mathsf{f}}(\boldsymbol{\sigma},\mathbf{b})=\mathbf{0}$ \citep{Morgan1966,Rajagopal2003,Rajagopal2007}. Cauchy elasticity is a subset of this class of solids.}
 In terms of the first Piola-Kirchhoff stress \citep{Truesdell1952,TruesdellNoll2004,Ogden1984}
\begin{equation}
	\mathbf{P}=\hat{\mathbf{P}}(X,\mathbf{F},\mathbf{G},\mathbf{g})\,.
\end{equation}
One can show that objectivity implies that the second Piola-Kirchhoff stress has the following functional form \citep{TruesdellNoll2004}
\begin{equation}
	\mathbf{S}=\hat{\mathbf{S}}(X,\mathbf{C}^\flat,\mathbf{G})\,.
\end{equation}
For an isotropic solid one has the following classic representation \citep{RivlinEricksen1955,Wang1969,Boehler1977}
\begin{equation}
	\mathbf{S}=\Lambda_0 \mathbf{G}^\sharp+\Lambda_1 \mathbf{C}^\sharp
	+\Lambda_{-1} \mathbf{C}^{-\sharp}\,,
\end{equation}
where $\Lambda_i=\Lambda_i(X,I_1,I_2,I_3)$, $i=-1,0,1$, and $\sharp$ is the sharp operator induced from the metric $\mathbf{G}$ (which raises indices). For incompressible isotropic Cauchy solids $I_3=1$, and one has
\begin{equation}
	\mathbf{S}=-p\,\mathbf{C}^{-\sharp}+\Lambda_0 \mathbf{G}^\sharp+\Lambda_1 \mathbf{C}^\sharp\,,
\end{equation}
where $p=p(X,t)$ is the Lagrange multiplier associated with the incompressibility constraint $J=\sqrt{I_3}=1$, and $\Lambda_i=\Lambda_i(X,I_1,I_2)$, $i=0,1$.

In terms of the Cauchy stress, the constitutive equations of compressible isotropic Cauchy elastic solids are written as
\begin{equation}
	\boldsymbol{\sigma} =\alpha\, \mathbf{g}^\sharp+\beta \mathbf{b}^\sharp
	+\gamma\mathbf{c}^\sharp\,, 
\end{equation}
where $\alpha=\alpha(X,I_1,I_2,I_3)$, $\beta=\beta(X,I_1,I_2,I_3)$, and $\gamma=\gamma(X,I_1,I_2,I_3)$ are arbitrary response functions. Similarly, for incompressible isotropic Cauchy elastic solids one has
\begin{equation}
\begin{aligned}
	\boldsymbol{\sigma} &= -p\, \mathbf{g}^\sharp+\beta \mathbf{b}^\sharp
	+\gamma\mathbf{c}^\sharp\,, 
\end{aligned}
\end{equation}
where $\beta=\beta(X,I_1,I_2)$ and $\gamma=\gamma(X,I_1,I_2)$ are arbitrary response functions.
In components, they read $\sigma^{ab}=\alpha\,g^{ab}+\beta\,b^{ab}+\gamma\,c^{ab}$ and $\sigma^{ab}=-p\,g^{ab}+\beta\,b^{ab}+\gamma\,c^{ab}$, respectively.

\citet{GreenNaghdi1971} showed that Cauchy elasticity is consistent with the first and second laws of thermodynamics. They demonstrated that over a closed path in the space of strains the net work in a Cauchy elastic solid may not vanish. They explicitly demonstrated this in the case of a linear stress-strain relationship and observed that the lack of major symmetries is responsible for the nonzero work.\footnote{The same thing was shown fifty years later in \citep{Scheibner2020} when formulating the so-called ``odd" elasticity. These authors  clearly were not aware of similar developments in the literature of nonlinear elasticity.}
There is no consensus in the literature of nonlinear elasticity on Cauchy elasticity being a viable theory; while some dismiss it \citep{Rivlin1986,Casey2005,Carroll2009,Rajagopal2011}, others appear to accept it \citep{Truesdell1952,Ericksen1956,TruesdellNoll2004,Ogden1984,Kruvzik2019,Bordiga2022}.
Our motivation for studying Cauchy elasticity is because of its promise in describing the mechanics of active matter  at large strains. Active matter may have access to external sources of energy, and hence, the net work done in a closed loop in the strain space may not vanish.
The recent interest in the physics literature in the so-called ``odd" elasticity \citep{Scheibner2020,Fruchart2023}, which is simply linearized non-hyperelastic Cauchy elasticity, is another motivation for revisiting Cauchy elasticity.

\section{Universal Deformations in Compressible Isotropic Cauchy Elasticity} \label{C-Elasticity-UC}

Let us consider a compressible isotropic Cauchy elastic body deforming in the Euclidean ambient space. In a local coordinate chart $\{x^a\}$, which may be curvilinear, the Cauchy stress has the following representation
\begin{equation}
	\sigma^{ab}=\alpha\,g^{ab}+\beta\,b^{ab}+\gamma\,c^{ab}\,. 
\end{equation}
When there are no body forces present, the equilibrium equations read $\sigma^{ab}{}_{|b}=0$.\footnote{$(.)_{|a}$ denotes covariant derivative with respect to the vector $\frac{\partial}{\partial x^a}$. In Cartesian coordinates, this reduces to a partial derivative. Also, it should be noted that for any scalar fiend $f$, $f_{|a}=f_{,a}$.} 
Thus\footnote{We have used the fact that a Riemannian metric is compatible with its Levi-Civita connection, i.e., $g^{ab}{}_{|c}=0$, and hence $g^{ab}{}_{|b}=0$, where summation over repeated indices is implied.}
\begin{equation} \label{Equilibrium-Compressible}
	\sigma^{ab}{}_{|b}=
	\beta\,b^{ab}{}_{|b}+\gamma\,c^{ab}{}_{|b}+
	\alpha_{,b}\,g^{ab}+\beta_{,b}\,b^{ab}+\gamma_{,b}\,c^{ab}
	=0\,. 
\end{equation}
Note that
\begin{equation}
\begin{aligned}
	&\alpha_{,b}=F^{-A}{}_b\,\frac{\partial \alpha}{\partial X^A}
	+\frac{\partial \alpha}{\partial I_1}I_{1,b}
	+\frac{\partial \alpha}{\partial I_2}I_{2,b}
	+\frac{\partial \alpha}{\partial I_3}I_{3,b}\,,\\
	& \beta_{,b}=F^{-A}{}_b\,\frac{\partial \beta}{\partial X^A}
	+\frac{\partial \beta}{\partial I_1}I_{1,b}
	+\frac{\partial \beta}{\partial I_2}I_{2,b}
	+\frac{\partial \beta}{\partial I_3}I_{3,b}\,,\\
	&\gamma_{,b}=F^{-A}{}_b\,\frac{\partial \gamma}{\partial X^A}
	+\frac{\partial \gamma}{\partial I_1}I_{1,b}
	+\frac{\partial \gamma}{\partial I_2}I_{2,b}
	+\frac{\partial \gamma}{\partial I_3}I_{3,b} \,.
\end{aligned}
\end{equation}
These can be written more concisely as
\begin{equation} \label{Coefficients-Derivatives}
\begin{aligned}
	&\alpha_{,b}=F^{-A}{}_b\,\alpha_{,A}+\alpha_1\,I_{1,b}+\alpha_2\,I_{2,b}+\alpha_3\,I_{3,b}\,,\\
	&\beta_{,b}=F^{-A}{}_b\,\beta_{,A}+\beta_1\,I_{1,b}+\beta_2\,I_{2,b}+\beta_3\,I_{3,b}\,,\\
	&\gamma_{,b}=F^{-A}{}_b\,\gamma_{,A}+\gamma_1\,I_{1,b}+\gamma_2\,I_{2,b}+\gamma_3\,I_{3,b} \,,
\end{aligned}
\end{equation}
where
\begin{equation} 
\begin{aligned}
	&\alpha_{,A}=\frac{\partial \alpha}{\partial X^A}\,, &&
	\beta_{,A}=\frac{\partial \beta}{\partial X^A}\,,&&
	\gamma_{,A}=\frac{\partial \gamma}{\partial X^A}\,,&& A=1,2,3\,,\\
	& \alpha_{i}=\frac{\partial \alpha}{\partial I_i}\,, &&
	\beta_{i}=\frac{\partial \beta}{\partial I_i}\,, &&
	\gamma_{i}=\frac{\partial \gamma}{\partial I_i}\,,&& i=1,2,3
	\,.
\end{aligned}
\end{equation}  
Substituting \eqref{Coefficients-Derivatives} into \eqref{Equilibrium-Compressible} one obtains
\begin{equation} 
\begin{aligned}
	&\beta\,b^{ab}{}_{|b}+\gamma\,c^{ab}{}_{|b} \\
	& + I_{1,b}\,g^{ab}\,\alpha_1+I_{2,b}\,g^{ab}\,\alpha_2+I_{3,b}\,g^{ab}\,\alpha_3 \\
	& + I_{1,b}\,b^{ab}\,\beta_1+I_{2,b}\,b^{ab}\,\beta_2+I_{3,b}\,b^{ab}\,\beta_3 \\
	& + I_{1,b}\,c^{ab}\,\gamma_1+I_{2,b}\,c^{ab}\,\gamma_2+I_{3,b}\,c^{ab}\,\gamma_3 \\
	& +F^{-A}{}_b\,\delta^{ab}\,\alpha_{,A}+F^{-A}{}_b\,b^{ab}\,\beta_{,A}	+F^{-A}{}_b\,c^{ab}\,\gamma_{,A}
	=0\,. 
\end{aligned}
\end{equation}
Note that $\alpha$, $\beta$, and $\gamma$ are arbitrary functions and their derivatives are independent. Therefore, for the equilibrium equations to hold for any compressible isotropic Cauchy elastic solid, the coefficient of each derivative must vanish. Thus
\begin{empheq}[left={\empheqlbrace }]{align}
	\label{Universality-Comp-1} 
	& b^{ab}{}_{|b}=c^{ab}{}_{|b}=0 \,,\\
	\label{Universality-Comp-2} 
	& g^{ab}I_{1,b}=g^{ab}I_{2,b}=g^{ab} I_{3,b}=0 \,, \\
	\label{Universality-Comp-3} 
	& b^{ab} I_{1,b}=b^{ab} I_{2,b}=b^{ab} I_{3,b}=0 \,, \\
	\label{Universality-Comp-4} 
	& c^{ab} I_{1,b}=c^{ab} I_{2,b}=c^{ab} I_{3,b}=0\,,\\
	\label{Universality-Comp-5} 
	& F^{-A}{}_b\,\delta^{ab}\,\alpha_{,A}=F^{-A}{}_b\,b^{ab}\,\beta_{,A}=F^{-A}{}_b\,c^{ab}\,\gamma_{,A}
	=0\,.
\end{empheq}

The universality constraints \eqref{Universality-Comp-2} imply that $I_1, I_2, I_3$ are constant. Then, the universality constraints \eqref{Universality-Comp-3} and \eqref{Universality-Comp-4} are trivially satisfied. From \eqref{Universality-Comp-5}, one concludes that $\alpha_{,A}=\beta_{,A}=\gamma_{,A}=0$, i.e., the body must be homogeneous. Thus, in summary we have concluded that
\begin{equation}
	I_1, I_2, I_3~\text{are constant},\quad\text{and}\qquad b^{ab}{}_{|b}=c^{ab}{}_{|b}=0\,.
\end{equation}
Using these and the compatibility equations, one can show that the universal deformations must be homogeneous \citep{Ericksen1955}. Thus, we have proved the following two results.

\begin{prop}
The set of universal deformations of homogeneous compressible isotropic Cauchy elastic solids is the set of all homogeneous deformations. 
\end{prop}

\begin{prop}
Inhomogeneous compressible isotropic Cauchy elastic solids do not admit universal deformations. 
\end{prop}

\begin{remark}
We observe that the set of universal deformations of homogeneous compressible isotropic Cauchy elastic solids is identical to that of homogeneous compressible isotropic hyperelastic solids \citep{Ericksen1955}. 
Additionally, it is noteworthy that neither inhomogeneous compressible isotropic Cauchy elastic solids nor inhomogeneous compressible isotropic hyperelastic solids admit universal deformations \citep{Yavari2021}.
\end{remark}

\section{Universal Deformations in Incompressible Isotropic Cauchy Elasticity} \label{C-Elasticity-UI}

For an incompressible isotropic Cauchy elastic solid the Cauchy stress has the following component representation 
\begin{equation}\label{Stress-Strain}
    \sigma^{ab}=-p\,g^{ab}+\beta\,b^{ab}+\gamma\,c^{ab}\,,
\end{equation}
where a curvilinear coordinate chart $\{x^a\}$ is assumed. Equilibrium equations in the absence of body forces read $\sigma^{ab}{}_{|b}=\sigma^{ab}{}_{,b}+\gamma^a{}_{bc}\,\sigma^{cb}+\gamma^b{}_{bc}\,\sigma^{ac}=0$, where $\gamma^a{}_{bc}$ (not to be confused with the scalar field $\gamma$ in \eqref{Stress-Strain}) are the Levi-Civita connection components corresponding to the metric $\mathbf{g}$.
Thus
\begin{equation}
	p_{,n}\,g^{an}=\beta\,b^{ab}{}_{|n}+\gamma\,c^{ab}{}_{|n}+\beta_{,n}\,b^{ab}+\gamma_{,n}\,c^{ab}\,,
\end{equation}
or\footnote{Note that $b^n{}_a=b^{nm}g_{ma}$, and $b_a{}^n=g_{am}b^{mn}$, which are equal. Thus, we use $b^n_a=b^n{}_a=b_a{}^n$. Similarly, the same notation is used for $\mathbf{c}$.}
\begin{equation}
	p_{,a}=\beta\,b^{n}{}_{a|n}+\gamma\,c^{n}_{a|n}+\beta_{,n}\,b^n_a+\gamma_{,n}\,c^n_a	\,.
\end{equation}
Thus, $\mathrm{d}p=p_{,a}\,\mathrm{d}x^a=(\beta\,b^{n}_{a|n}+\gamma\,c^{n}_{a|n}+\beta_{,n}\,b^n_a+\gamma_{,n}\,c^n_a)\,dx^a=\boldsymbol{\xi}$, where $\mathrm{d}$ is the exterior derivative. This means the one-form $\boldsymbol{\xi}$ must be exact. A necessary condition for $\boldsymbol{\xi}$ to be exact is $\mathrm{d}\boldsymbol{\xi}=0$. This implies that $p_{,ab}=p_{,ba}$, which is equivalent to $p_{|ab}=p_{|ba}$. Thus
\begin{equation}
	p_{|ab}=\beta\,b^{n}_{a|nb}+\gamma\,c^{n}_{a|nb}+\beta_{,n}\,b^n_{a|b}+\gamma_{,n}\,c^n_{a|b}
	+\beta_{,b}\,b^{n}_{a|n}+\gamma_{,b}\,c^{n}_{a|n}+(\beta_{,n})_{|b}\,b^n_a+(\gamma_{,n})_{|b}\,c^n_a
	\,,
\end{equation}
must be symmetric in $(a,b)$.

Note that
\begin{equation}
	\beta_{,n}= F^{-A}{}_n\,\beta_{,A}+\beta_1\,I_{1,n}+\beta_2\,I_{2,n} \,,\qquad
	\gamma_{,n}=F^{-A}{}_n\,\gamma_{,A}+\gamma_1\,I_{1,n}+\gamma_2\,I_{2,n} \,,
\end{equation}
where 
\begin{equation}
\begin{aligned}
	& \beta_{i}=\frac{\partial \beta}{\partial I_i} \,,&&
	\gamma_{i}=\frac{\partial \gamma}{\partial I_i}\,,&& i=1,2 \,,\\
	& \beta_{,A}=\frac{\partial \beta}{\partial X^A} \,,&&
	\gamma_{,A}=\frac{\partial \gamma}{\partial X^A}\,,&& A=1,2,3\,.
\end{aligned}
\end{equation}
Also
\begin{equation}
\begin{aligned}
	(\beta_{,n})_{|b} &=\beta_1\,I_{1|nb}+\beta_2\,I_{2|nb}+\beta_{1,b}\,I_{1,n}+\beta_{2,b}\,I_{2,n}
	+\left( F^{-B}{}_b\,F^{-A}{}_{n,B}-\gamma^m{}_{nb}\,F^{-A}{}_m \right) \beta_{,A}\\
	& \quad+F^{-A}{}_n\left(F^{-B}{}_b\,\beta_{,AB}+\beta_{1,A}I_{1,b}+\beta_{2,A}I_{2,b} \right) \,,\\
	&= \beta_1\,I_{1|nb}+\beta_2\,I_{2|nb}
	+I_{1,b}\,I_{1,n}\,\beta_{11}+I_{2,b}\,I_{2,n}\,,\beta_{22}
	+\left(I_{1,b}\,I_{2,n}+I_{1,n}\,I_{2,b}\right)\beta_{12} \\
	&\quad +\left( F^{-B}{}_b\,F^{-A}{}_{n,B}-\gamma^m{}_{nb}\,F^{-A}{}_m \right) \beta_{,A}
	+\frac{1}{2}\left(F^{-A}{}_n\,F^{-B}{}_b+F^{-B}{}_n\,F^{-A}{}_b\right)\beta_{,AB} \\
	& \quad + \left(F^{-A}{}_b\,I_{1,n}+F^{-A}{}_n\,I_{1,b}\right)\beta_{1,A}
	+\left(F^{-A}{}_b\,I_{2,n}+F^{-A}{}_n\,I_{2,b}\right)\beta_{2,A}  \,, 
\end{aligned}
\end{equation}
and
\begin{equation}
\begin{aligned}
	(\gamma_{,n})_{|b} &=\gamma_1\,I_{1|nb}+\gamma_2\,I_{2|nb}
	+\gamma_{1,b}\,I_{1,n}+\gamma_{2,b}\,I_{2,n}
	+\left( F^{-B}{}_b\,F^{-A}{}_{n,B}-\gamma^m{}_{nb}\,F^{-A}{}_m \right) \gamma_{,A}\\
	& \quad+F^{-A}{}_n\left(F^{-B}{}_b\,\gamma_{,AB}+\gamma_{1,A}I_{1,b}+\gamma_{2,A}I_{2,b} \right) \,,\\
	&= \gamma_1\,I_{1|nb}+\gamma_2\,I_{2|nb}
	+I_{1,b}\,I_{1,n}\,\gamma_{11}+I_{2,b}\,I_{2,n}\,,\gamma_{22}
	+\left(I_{1,b}\,I_{2,n}+I_{1,n}\,I_{2,b}\right)\gamma_{12} \\
	&\quad +\left( F^{-B}{}_b\,F^{-A}{}_{n,B}-\gamma^m{}_{nb}\,F^{-A}{}_m \right) \gamma_{,A}
	+\frac{1}{2}\left(F^{-A}{}_n\,F^{-B}{}_b+F^{-B}{}_n\,F^{-A}{}_b\right)\gamma_{,AB} \\
	& \quad + \left(F^{-A}{}_b\,I_{1,n}+F^{-A}{}_n\,I_{1,b}\right)\gamma_{1,A}
	+\left(F^{-A}{}_b\,I_{2,n}+F^{-A}{}_n\,I_{2,b}\right)\gamma_{2,A}  \,, 
\end{aligned}
\end{equation}
where
\begin{equation}
\begin{aligned}
	& \beta_{ij}=\frac{\partial^2 \beta}{\partial I_i\,\partial I_j} \,, &&
	\gamma_{ij}=\frac{\partial^2 \gamma}{\partial I_i\,\partial I_j}\,,&& i \leq j=1,2\,, \\
	& \beta_{i,A}=\frac{\partial^2 \beta}{\partial I_i\,\partial X^A} \,,&&
	\gamma_{i,A}=\frac{\partial^2 \gamma}{\partial I_i\,\partial X^A}\,,&& i=1,2\,,~A=1,2,3\,,\\
	& \beta_{,AB}=\frac{\partial^2 \beta}{\partial X^A\,\partial X^B} \,,&& 
	\gamma_{,AB}=\frac{\partial^2 \gamma}{\partial X^A\,\partial X^B} \,,&& A \leq B=1,2,3
	\,.
\end{aligned}
\end{equation}
Therefore
\begin{equation}
\begin{aligned}
	p_{|ab} &= b^{n}{}_{a|nb}\,\beta+c^{n}_{a|nb}\,\gamma \\
	& +\left[ I_{1,b}\,b^n_{a|n}+(b^n_a\,I_{1,n})_{|b} \right]\beta_1
	+\left[ I_{2,b}\,b^n_{a|n}+(b^n_a\,I_{2,n})_{|b} \right]\beta_2   \\
	& +I_{1,n}\,I_{1,b}\,b^n_a\,\beta_{11}+I_{2,n}\,I_{2,b}\,b^n_a\,\beta_{22}
	+\left(I_{1,n}\,I_{2,b}+I_{2,n}\,I_{1,b}\right)b^n_a\,\beta_{12}  \\
	& +\left[ I_{1,b}\,c^n_{a|n}+(c^n_a\,I_{1,n})_{|b} \right]\gamma_1
	+\left[ I_{2,b}\,c^n_{a|n}+(c^n_a\,I_{2,n})_{|b} \right]\gamma_2 \quad  \\
	& +I_{1,n}\,I_{1,b}\,c^n_a\,\gamma_{11}+I_{2,n}\,I_{2,b}\,c^n_a\,\gamma_{22}
	+\left(I_{1,n}\,I_{2,b}+I_{2,n}\,I_{1,b}\right)c^n_a\,\gamma_{12} \\
	& +\left[F^{-A}{}_n\,b^n_{a|b}+F^{-A}{}_b\,b^n_{a|n}  
	+\left( F^{-B}{}_b\,F^{-A}{}_{n,B}-\gamma^m{}_{nb}\,F^{-A}{}_m \right) b^n_a\right] \beta_{,A}  \\
	& +\left[F^{-A}{}_n\,c^n_{a|b}+F^{-A}{}_b\,c^n_{a|n}  
	+\left( F^{-B}{}_b\,F^{-A}{}_{n,B}-\gamma^m{}_{nb}\,F^{-A}{}_m \right) c^n_a\right] \gamma_{,A}   \\
	& +F^{-A}{}_n\,b^n_a\,I_{1,b}\,\beta_{1,A}+F^{-A}{}_n\,b^n_a\,I_{2,b}\,\beta_{2,A}
	    +F^{-A}{}_n\,c^n_a\,I_{1,b}\,\gamma_{1,A}+F^{-A}{}_n\,c^n_a\,I_{2,b}\,\gamma_{2,A} \\
	& +F^{-A}{}_n\,F^{-B}{}_b\,b^n_a\,\beta_{,AB}+F^{-A}{}_n\,F^{-B}{}_b\,c^n_a\,\gamma_{,AB}
	\,.
\end{aligned}
\end{equation}
The functions $\beta$ and $\gamma$ and their derivatives are independent, and hence, the coefficient of each function must be symmetric. Thus, we have the following set of universality constraints for homogeneous Cauchy elasticity
\begin{empheq}[left={\empheqlbrace }]{align}
	\label{Universality-1} 
	\mathscr{A}^0_{ab} &= \textcolor{blue}{b^{n}{}_{a|nb} } \,,\\
	\label{Universality-2} 
	\mathscr{A}^1_{ab} &= \textcolor{blue}{I_{1,b}\,b^n_{a|n}+(b^n_a\,I_{1,n})_{|b}} \,, \\
	\label{Universality-3} 
	\mathscr{A}^2_{ab} &= I_{2,b}\,b^n_{a|n}+(b^n_a\,I_{2,n})_{|b} \,, \\
	\label{Universality-4} 
	\mathscr{A}^{11}_{ab} &= \textcolor{blue}{I_{1,n}\,I_{1,b}\,b^n_a}  \,,\\
	\label{Universality-5} 
	\mathscr{A}^{22}_{ab} &= I_{2,n}\,I_{2,b}\,b^n_a \,,\\
	\label{Universality-6} 
	\mathscr{A}^{12}_{ab} &= \left(I_{1,n}\,I_{2,b}+I_{2,n}\,I_{1,b}\right)b^n_a \,, \\
	\label{Universality-7} 
	\mathscr{B}^0_{ab} &= \textcolor{blue}{c^{n}_{a|nb}}  \,,\\
	\label{Universality-8} 
	\mathscr{B}^1_{ab} &= I_{1,b}\,c^n_{a|n}+(c^n_a\,I_{1,n})_{|b}  \,, \\
	\label{Universality-9} 
	\mathscr{B}^2_{ab} &= \textcolor{blue}{I_{2,b}\,c^n_{a|n}+(c^n_a\,I_{2,n})_{|b}}  \,, \\
	\label{Universality-10} 
	\mathscr{B}^{11}_{ab} &= I_{1,n}\,I_{1,b}\,c^n_a \,,\\
	\label{Universality-11} 
	\mathscr{B}^{22}_{ab} &= \textcolor{blue}{I_{2,n}\,I_{2,b}\,c^n_a} \,,\\
	\label{Universality-12} 
	\mathscr{B}^{12}_{ab} &= \left(I_{1,n}\,I_{2,b}+I_{2,n}\,I_{1,b}\right)c^n_a
	\,.
\end{empheq}
The six terms highlighted in blue are identical to those of hyperealsticity \citep{Ericksen1954}.
The following is the set of universal inhomogeneity constraints inhomogeneous Cauchy elasticity
\begin{empheq}[left={\empheqlbrace }]{align}
	\label{IUniversality-1} 
	\mathscr{C}^{0A}_{ab} &= \textcolor{purple}{F^{-A}{}_n\,b^n_{a|b}+F^{-A}{}_b\,b^n_{a|n}  
	+\left( F^{-B}{}_b\,F^{-A}{}_{n,B}-\gamma^m{}_{nb}\,F^{-A}{}_m \right) b^n_a}  \,,\\
	\label{IUniversality-2} 
	\mathscr{C}^{1A}_{ab} &= \textcolor{purple}{\left(F^{-A}{}_n\,I_{1,b}+F^{-A}{}_b\,I_{1,n} \right)b^n_a} \,, \\
	\label{IUniversality-3} 
	\mathscr{C}^{2A}_{ab} &= \left(F^{-A}{}_n\,I_{2,b}+F^{-A}{}_b\,I_{2,n} \right) b^n_a \,, \\
	\label{IUniversality-4} 
	\mathscr{C}^{0AB}_{ab} &=  \textcolor{purple}{\left(F^{-A}{}_n\,F^{-B}{}_b+F^{-B}{}_n\,F^{-A}{}_b\right)  
	b^n_a}  \,,\\
	\label{IUniversality-5} 
	\mathscr{D}^{0A}_{ab} &= \textcolor{purple}{F^{-A}{}_n\,c^n_{a|b}+F^{-A}{}_b\,c^n_{a|n}  
	+\left( F^{-B}{}_b\,F^{-A}{}_{n,B}-\gamma^m{}_{nb}\,F^{-A}{}_m \right) c^n_a}  \,,\\
	\label{IUniversality-6} 
	\mathscr{D}^{1A}_{ab} &= \left(F^{-A}{}_n\,I_{1,b}+F^{-A}{}_b\,I_{1,n} \right)c^n_a  \,, \\
	\label{IUniversality-7} 
	\mathscr{D}^{2A}_{ab} &= \textcolor{purple}{\left(F^{-A}{}_n\,I_{2,b}+F^{-A}{}_b\,I_{2,n} \right) c^n_a} \,, \\
	\label{IUniversality-8} 
	\mathscr{D}^{0AB}_{ab} &=  \textcolor{purple}{\left(F^{-A}{}_n\,F^{-B}{}_b+F^{-B}{}_n\,F^{-A}{}_b\right) 
	c^n_a}
	\,.
\end{empheq}
These terms must be symmetric in $(a,b)$ for $A=1,2,3$ and $A\leq B=1,2,3$. The six terms highlighted in purple are identical to those of inhomogeneous isotropic hyperelasticity \citep{Yavari2021}.

Recall that Ericksen's universality constraints of isotropic hyperelasticity are \citep{Ericksen1954}:
\begin{empheq}[left={\empheqlbrace }]{align} 
	\label{Universality-E1} 
	\mathcal{A}_{ab}^{1} & = \textcolor{blue}{b_a^{n}{}_{|bn}}\,, \\
	\label{Universality-E2} 
	\mathcal{A}_{ab}^{2} &= \textcolor{blue}{c_a^{n}{}_{|bn}} \,, \\
	\label{Universality-E3} 
	\mathcal{A}_{ab}^{11} &= \textcolor{blue}{b_a^{n}{}_{|n}\,I_{1,b}+\left(b_a^n\,I_{1,n}\right)_{|b}}\,,  \\
	\label{Universality-E4} 
	\mathcal{A}_{ab}^{22} &= \textcolor{blue}{c_a^{n}{}_{|n}\,I_{2,b}+\left(c_a^{n}\,I_{2,n}\right)_{|b}}\,, \\
	\label{Universality-E5} 
	\mathcal{A}_{ab}^{12} &= \left(b_a^{n}\,I_{2,n}\right)_{|b}+b_a^{n}{}_{|n}\,I_{2,b}
	-\left[\left(c_a^{n}\,I_{1,n}\right)_{|b}+c_a^{n}{}_{|n}\,I_{1,b}  \right]\,,\\
	\label{Universality-E6} 
	\mathcal{A}_{ab}^{111} &= \textcolor{blue}{b_a^{n}\,I_{1,n}I_{1,b}}\,, \\
	\label{Universality-E7} 
	\mathcal{A}_{ab}^{222} &= \textcolor{blue}{c_a^{n}\,I_{2,n}I_{2,b}} \,, \\
	\label{Universality-E8} 
	\mathcal{A}_{ab}^{112} &= b_a^{n}\left(I_{1,b}I_{2,n}+I_{1,n}I_{2,b}\right)-c_a^{n}\,I_{1,n}I_{1,b}\,, \\
	\label{Universality-E9} 
	\mathcal{A}_{ab}^{122} &= b_a^{n}\,I_{2,b}I_{2,n}-c_a^{n}\left(I_{1,b}I_{2,n}+I_{1,n}I_{2,b}\right)\,.
\end{empheq}
Also the universal inhomogeneity constraints of hyperelasticity are \citep{Yavari2021}:
\begin{empheq}[left={\empheqlbrace }]{align}
	\label{IUniversality-H1} 
	\mathcal{C}_{ab}^{1A} & = \textcolor{purple}{F^{-A}{}_n\,b_a^n{}_{|b} + F^{-A}{}_b\,b_a^n{}_{|n}
	+b_a^n\left[F^{-B}{}_b\,F^{-A}{}_{n,B}-\gamma^m{}_{nb}\,F^{-A}{}_m \right]} \,, \\
	\label{IUniversality-H2} 
	\mathcal{C}_{ab}^{2A} & = \textcolor{purple}{F^{-A}{}_n\,c_a^n{}_{|b} + F^{-A}{}_b\,c_a^n{}_{|n}
	+c_a^n\left[F^{-B}{}_b\,F^{-A}{}_{n,B}-\gamma^m{}_{nb}\,F^{-A}{}_m \right]} \,, \\
	\label{IUniversality-H3} 
	\mathcal{C}_{ab}^{11A} & =\textcolor{purple}{b_a^n\left[F^{-A}{}_n\,I_{1,b}+F^{-A}{}_b\,I_{1,n} \right]} \,, \\
	\label{IUniversality-H4} 
	\mathcal{C}_{ab}^{22A} &= \textcolor{purple}{c_a^n\left[F^{-A}{}_n\,I_{2,b}+F^{-A}{}_b\,I_{2,n} \right]} \,, \\
	\label{IUniversality-H5} 
	\mathcal{C}_{ab}^{12A} & = b_a^n\left[F^{-A}{}_n\,I_{2,b}+F^{-A}{}_b\,I_{2,n} \right]
	-c_a^n\left[F^{-A}{}_n\,I_{1,b}+F^{-A}{}_b\,I_{1,n} \right] \,, \\	
	\label{IUniversality-H6} 
	\mathcal{C}_{ab}^{1AB} & =\textcolor{purple}{b^n_a \left[F^{-A}{}_n\,F^{-B}{}_b
	+(F^{-1})^B{}_n\,F^{-A}{}_b \right] }	\,, \\
	\label{IUniversality-H7} 
	\mathcal{C}_{ab}^{2AB} & = \textcolor{purple}{c^n_a \left[F^{-A}{}_n\,F^{-B}{}_b
	+(F^{-1})^B{}_n\,F^{-A}{}_b \right]}  \,.
\end{empheq}

First, notice that there are nine sets of universality constraints \eqref{Universality-E1}-\eqref{Universality-E9} in homogeneous isotropic hyperelasticity \citep{Ericksen1954} compared to twelve in compressible isotropic Cauchy elasticity \eqref{Universality-1}-\eqref{Universality-12}. Second, the six terms shown in blue are identical to those of hyperelasticity. Thus, we first look at these six terms.

\citet{Ericksen1954} used the following result. Suppose $\mathbf{u}$ and $\mathbf{v}$ are vectors such that $\mathbf{u}\otimes\mathbf{v}=\mathbf{v}\otimes\mathbf{u}$. Assuming that $\mathbf{v}\neq\mathbf{0}$ (if both vectors are zero this equality trivially holds) one can write
\begin{equation}
	\mathbf{u}=\frac{\mathbf{u}\cdot\mathbf{v}}{|\mathbf{v}|^2}\mathbf{v}=\lambda\mathbf{v},\,
\end{equation}
i.e., $\mathbf{u}$ and $\mathbf{v}$ must be parallel.
Now suppose that $\mathbf{u}^{\flat}=\mathrm{d}\phi$ and $\mathbf{v}^{\flat}=\mathrm{d}\psi$, where $\flat$ is the flat operator that gives the $1$-form corresponding to a vector, $\mathrm{d}$ is the exterior derivative, and $\phi$ and $\psi$ are scalar fields. Note that $\lambda=\lambda(\mathbf{u},\mathbf{v})=\lambda(\phi,\psi)$. Thus, $\mathbf{u}^{\flat}=\mathrm{d}\phi=\lambda(\phi,\psi)\mathrm{d}\psi$. Hence
\begin{equation}
	0=\mathrm{d}\circ\mathrm{d}\phi=\mathrm{d}\lambda(\phi,\psi)\wedge\mathrm{d}\psi
	=\frac{\partial \lambda}{\partial \phi}\,\mathrm{d}\phi\wedge\mathrm{d}\psi\,,
\end{equation}
where $\wedge$ is the wedge product of differential forms. Therefore, $\lambda=\lambda(\psi)$, i.e., $\mathbf{u}^{\flat}=\lambda(\psi)\mathrm{d}\psi$.

\paragraph{$W_{111}$ ($\mathscr{A}^{11}_{ab}$) and $W_{222}$ $(\mathscr{B}^{22}_{ab})$ terms:}
Symmetry of the coefficient of the $\mathscr{A}^{11}_{ab}$ term implies that \citep{Ericksen1954}
\begin{equation}
	(b^n_aI_{1,n})I_{1,b}=(b^n_bI_{1,n})I_{1,a}\,.
\end{equation}
One concludes that either $I_{1,a}=0$ or $b^n_a\,I_{1,n}=\mathsf{A}^{-1}\,I_{1,a}$, for some scalar function $\mathsf{A}$. This means that $I_{1,a}$ is an eigenvector of $b^b_a$ with eigenvalue $\mathsf{A}^{-1}$. We know that $b^{am}c_{mb}=\delta^a_b$, and hence, $b^m_bc^a_m=\delta_b^a$. Similarly, $b_{am}c^{mb}=\delta_a^b$, and hence, $b^a_mc^m_b=\delta^a_b$. Therefore, $c^n_a\,I_{1,n}=\mathsf{A}\,I_{1,a}$.\footnote{Note that both $\mathbf{b}$ and $\mathbf{c}$ are positive-definite. This means that when $I_1$ is not constant, $\mathsf{A}\neq 0$, and hence, $\mathsf{A}^{-1}$ is an eigenvalue of $\mathbf{c}$.}
Similarly, the symmetry of the coefficient of the $\mathscr{B}^{22}_{ab}$ term implies that
\begin{equation}
	(c^n_aI_{2,n})I_{2,b}=(c^n_bI_{2,n})I_{2,a}\,.
\end{equation}
The conclusion is that either $I_{2,a}=0$ or $c^n_a\,I_{2,n}=\bar{\mathsf{A}}\,I_{2,a}$, for some scalar function $\bar{\mathsf{A}}$. One also has $b^n_a\,I_{2,n}=\bar{\mathsf{A}}^{-1}\,I_{2,a}$. In summary
\begin{empheq}[left={\empheqlbrace }]{align} \label{Eigenvalues-b-c}
	& c^n_a\,I_{1,n}=\mathsf{A}\,I_{1,a}  \,,\\
	& c^n_a\,I_{2,n}=\bar{\mathsf{A}}\,I_{2,a} \,,\\
	& b^n_a\,I_{1,n}=\mathsf{A}^{-1}\,I_{1,a}  \,,\\
	& b^n_a\,I_{2,n}=\bar{\mathsf{A}}^{-1}\,I_{2,a}
	\,.
\end{empheq}

\paragraph{$\mathscr{A}^{22}_{ab}$ and $\mathscr{B}^{11}_{ab}$ terms:}Symmetry of the coefficient of the $\mathscr{A}^{22}_{ab}$ term implies that
\begin{equation}
	(b^n_aI_{2,n})I_{2,b}=(b^n_bI_{2,n})I_{2,a}\,,\quad\text{or}\qquad
	\bar{\mathsf{A}}^{-1}I_{2,a}\,I_{2,b}=\bar{\mathsf{A}}^{-1}I_{2,b}\,I_{2,a}
	\,,
\end{equation}
which is trivially satisfied.
Similarly, the symmetry of the coefficient of the $\mathscr{B}^{11}_{ab}$ term implies that
\begin{equation}
	(c^n_aI_{1,n})I_{1,b}=(c^n_bI_{1,n})I_{1,a}\,,\quad\text{or}\qquad 
	\mathsf{A}\,I_{1,a}\,I_{1,b}=\mathsf{A}\,I_{1,b}\,I_{1,a}
	\,,
\end{equation}
which is also trivially satisfied.

\paragraph{$\mathscr{A}^{12}_{ab}$ and $\mathscr{B}^{12}_{ab}$ terms:}Symmetry of the coefficient of the $\mathscr{A}^{12}_{ab}$ term implies that
\begin{equation}
	 \left(I_{1,n}\,I_{2,b}+I_{2,n}\,I_{1,b}\right)b^n_a=\left(I_{1,n}\,I_{2,a}+I_{2,n}\,I_{1,a}\right)b^n_b
	\,,
\end{equation}
which is identical to the symmetry condition that comes from the coefficient of the $W_{112}$ in hyperelasticity \citep{Ericksen1954}. Thus
\begin{equation}
	(I_{1,b}I_{2,n}+I_{1,n}I_{2,b})b^n_a=(I_{1,a}I_{2,n}+I_{1,n}I_{2,a})b^n_b\,.
\end{equation}
Hence
\begin{equation}
	I_{1,b}\bar{\mathsf{A}}^{-1}I_{2,a}+\mathsf{A}^{-1}I_{1,a}I_{2,b}
	=I_{1,a}\bar{\mathsf{A}}^{-1}I_{2,b}+\mathsf{A}^{-1}I_{1,b}I_{2,a}\,.
\end{equation}
Or
\begin{equation}\label{A-A}
	\left(\mathsf{A}^{-1}-\bar{\mathsf{A}}^{-1}\right)(I_{1,a}I_{2,b}-I_{1,b}I_{2,a})=0\,.
\end{equation}
Similarly, the symmetry of the coefficient of the $\mathscr{B}^{12}_{ab}$ term implies that
\begin{equation}
	\left(I_{1,n}\,I_{2,b}+I_{2,n}\,I_{1,b}\right)c^n_a=\left(I_{1,n}\,I_{2,a}+I_{2,n}\,I_{1,a}\right)c^n_b
	\,.
\end{equation}
This is simplified to read
\begin{equation} \label{A-A1}
	\left(\mathsf{A}-\bar{\mathsf{A}}\right)(I_{1,a}I_{2,b}-I_{1,b}I_{2,a})=0\,,
\end{equation}
which is identical to the symmetry condition that comes from the coefficient of the $W_{221}$ in hyperelasticity \citep{Ericksen1954}. If either $I_1$ or $I_2$ is constant, \eqref{A-A} is satisfied for any $\mathsf{A}$ and $\bar{\mathsf{A}}$. If $\bar{\mathsf{A}}\neq \mathsf{A}$, one has $I_{1,a}I_{2,b}=I_{1,b}I_{2,a}$, which implies that $I_{1,a}$ and $I_{2,a}$ are parallel. However, these are eigenvectors of $b^b_a$ corresponding to the distinct eigenvalues $\bar{\mathsf{A}}\neq \mathsf{A}$ and must be perpendicular. This contradiction shows that $\bar{\mathsf{A}}= \mathsf{A}$.

Next, Ericksen shows that $I_1$ and $I_2$ must be functionally dependent.\footnote{$I_1$ and $I_2$ are functionally dependent if there exists a non-trivial function (a function that is not identically zero) such that $F(I_1,I_2)=0$. Taking derivatives one obtains
\begin{equation}\label{Jacobian}
\begin{bmatrix}
  I_{1,1} & I_{2,1}   \\
  I_{1,2} & I_{2,2}   \\
  I_{1,3} & I_{2,3}   
\end{bmatrix}
\begin{bmatrix} \frac{\partial F}{\partial I_1} \\ \frac{\partial F}{\partial I_2} \end{bmatrix}
=\begin{bmatrix} 0 \\ 0 \\0 \end{bmatrix}\,.
\end{equation}
For $F$ to be non-trivial the Jacobian matrix must have rank less that $2$.
} If either $I_1$ or $I_2$ is constant, this is trivially the case. If all the eigenvalues of $\mathbf{b}$ (and hence those of $\mathbf{c}$) are distinct, each must have a one-dimensional eigenspace. Therefore, $I_{1,a}$ and $I_{2,a}$, which are eigenvectors corresponding to the same eigenvalue, must be parallel. This guarantees that the rank of the Jacobian matrix in \eqref{Jacobian} is less that $2$, and hence, $I_1$ and $I_2$ are functionally dependent. Note that
\begin{equation}
	I_1=\frac{1}{c_1}+\frac{1}{c_2}+\frac{1}{c_3}\,,\qquad
	I_2=\frac{1}{c_1c_2}+\frac{1}{c_2c_3}+\frac{1}{c_3c_1}\,,\qquad
	c_1c_2c_3=1\,,
\end{equation}
where $c_i$ are eigenvalues of $\mathbf{c}$. If two eigenvalues are equal, e.g., $c_1=c_2$, one has (the case $c_1=c_2=c_3$ is trivial as incompressibility implies $c_1=c_2=c_3=1$.)
\begin{equation}
	I_1=\frac{2}{c_1}+c_1^2\,,\qquad I_2=\frac{1}{c_1^2}+2c_1\,.
\end{equation}
Clearly, in this case $I_1$ and $I_2$ are functionally dependent. Therefore, there is a scalar function $\mathsf{B}$ such that, $I_1=I_1(\mathsf{B})$ and $I_2=I_2(\mathsf{B})$. Similarly, $c_i=c_i(\mathsf{B})$. 

\paragraph{$W_{11}$ ($\mathscr{A}^1_{ab}$) and $W_{22}$ ($\mathscr{B}^2_{ab}$) terms:}Note that $I_{i,a}=I_i'(\mathsf{B})\mathsf{B}_{,a}$, $i=1,2$. The second term in the coefficient of $W_{22}$ ($\mathscr{B}^2_{ab}$) is simplified to read $(c^n_aI_{2,n})_{,b}=(c_1 I_2'\mathsf{B}_{,a})_{,b}=(c_1 I_2')'\mathsf{B}_{,b}\,\mathsf{B}_{,a}+c_1\, I_2'\,\mathsf{B}_{,ab}$, which is symmetric. Therefore, symmetry of the coefficient of $W_{22}$ implies that
\begin{equation}
	I_2'(c^n_{a|n}\,\mathsf{B}_{,b}-c^n_{b|n}\,\mathsf{B}_{,a})=0\,.
\end{equation}
Thus, either $I_2$ is constant or
\begin{equation}\label{D}
	c^n_{a,n}=\mathsf{D}\,\mathsf{B}_{,a}\,,
\end{equation}
where $\mathsf{D}$ is a scalar field. Similarly, symmetry of the coefficient of $W_{11}$ ($\mathscr{A}^1_{ab}$) dictates that either $I_1$ is constant or
\begin{equation}\label{E}
	b^n_{a|n}=\mathsf{E}\,\mathsf{B}_{,a}\,,
\end{equation}
where $\mathsf{E}$ is a scalar field.

\paragraph{$\mathscr{A}^2_{ab}$ and $\mathscr{B}^1_{ab}$ terms:}The symmetry of $\mathscr{A}^2_{ab}$ terms implies that
\begin{equation}
	I_2'(b^n_{a|n}\,\mathsf{B}_{,b}-b^n_{b|n}\,\mathsf{B}_{,a})
	=I_2'\left(\mathsf{E}\,\mathsf{B}_{,a}\,\mathsf{B}_{,b}-\mathsf{E}\,\mathsf{B}_{,b}\,\mathsf{B}_{,a}\right)=0\,,
\end{equation}
which is trivially satisfied.
Similarly, the symmetry of $\mathscr{B}^1_{ab}$ terms implies that
\begin{equation}
	I_1'\left(c^n_{a|n}\mathsf{B}_{,b}-c^n_{b|n}\mathsf{B}_{,a}\right)
	=I_1'\left(\mathsf{D}\mathsf{B}_{,a}\mathsf{B}_{,b}-\mathsf{D}\mathsf{B}_{,b}\mathsf{B}_{,a}\right)=0\,,
\end{equation}
which is also trivially satisfied.

The only two remaining universality constraints are $\mathscr{A}^0_{ab}$ and $\mathscr{B}^0_{ab}$ terms, which are identical to the coefficients of $W_1$ and $W_2$ in hyperelasticity. Using this two sets of universality constraints, \citet{Ericksen1954} found Families $1$-$4$ of universal deformations.

Note that if $I_1$ and $I_2$ are constant, \eqref{Universality-1}-\eqref{Universality-12} reduce to the symmetry of the two terms $b^{n}_{a|nb}$ and $c^{n}_{a|nb}$, which is exactly what one would have in hyperealsticity.
We have thus shown that although there are more universality constraints in incompressible isotropic Cauchy elasticity and some of the universality constraints are not identical to those of hyperelasticity, nevertheless the two sets of universality constraints are equivalent. 

Next we show that the universal inhomogeneity constraints of Cauchy elasticity \eqref{IUniversality-1}-\eqref{IUniversality-8} are equivalent to those of hyperelasticity \eqref{IUniversality-H1}-\eqref{IUniversality-H7}. We only need to check the terms in black, as the ones highlighted in purple are identical. Let us define $\mathsf{P}^A_{bn}=F^{-A}{}_n\,I_{1,b}+F^{-A}{}_b\,I_{1,n}$ and note that $\mathsf{P}^A_{nb}=\mathsf{P}^A_{bn}$. From \eqref{IUniversality-2} we know that
\begin{equation}
	\mathsf{P}^A_{bn}\,b^n_a = \mathsf{P}^A_{an}\,b^n_b \,.
\end{equation}
Multiply both sides by $c^a_m\,c^b_k$ to obtain $c^b_k\,\mathsf{P}^A_{bn}\,b^n_a\,c^a_m = c^a_m\,\mathsf{P}^A_{an}\,b^n_b\,c^b_k$, or $c^b_k\,\mathsf{P}^A_{bn}\,\delta^n_m = c^a_m\,\mathsf{P}^A_{an}\,\delta^n_k$.
Thus
\begin{equation}
	c^b_k\,\mathsf{P}^A_{bm} = c^a_m\,\mathsf{P}^A_{ak}\,, 
	\quad\text{or}\qquad
	c^n_a\,\mathsf{P}^A_{nb} = c^n_b\,\mathsf{P}^A_{na} \,,
\end{equation}
which is \eqref{IUniversality-6}. Similarly, \eqref{IUniversality-7} is equivalent to \eqref{IUniversality-3}. Moreover, symmetry of the terms \eqref{IUniversality-H3} and \eqref{IUniversality-H4} imply the symmetry of \eqref{IUniversality-H5}.
Therefore, we have proved the following result.

\begin{prop}
The set of universal deformations of incompressible isotropic Cauchy elasticity is identical to that of incompressible isotropic hyperelasticity. For a given universal deformation, the corresponding set of universal inhomogeneities is identical to that of hyperelasticity.
\end{prop}

\section{Conclusions}  \label{Sec:Conclusions}

The existing studies of universal deformations have been restricted to hyperelasticity, or more generally, to materials that have an underlying energy function. In this paper, we extended the analysis of universal deformations to Cauchy elastic solids, which in general, do not have an energy function.
We considered both compressible and incompressible inhomogeneous isotropic Cauchy elasticity. As hyperelasticity is a proper subset of Cauchy elasticity, one would expect the set of universal deformations of Cauchy elasticity to be a proper subset of that of hyperelasticity. 
We proved the somewhat unexpected result that the sets of universal deformations of isotropic Cauchy elasticity and isotropic hyperelasticity are identical, in both the compressible and incompressible cases.
We also proved that their corresponding universal inhomogeneities are identical as well.

\section*{Acknowledgments}

This work was supported by NSF -- Grant No. CMMI 1939901.

\bibliographystyle{abbrvnat}
\bibliography{ref,ref1}

\end{document}